
\input jnl
\input eqnorder
\input reforder
\def\tr{{\rm Tr}\,}
\preprintno{}
\title
Mean Area of Self-Avoiding Loops
\author
John Cardy
\affil
\def\oxon{All Souls College\\{\rm and}\\Department of Physics\\
Theoretical Physics\\1 Keble Road\\Oxford OX1 3NP, UK}
\oxon
\abstract
The mean area of two-dimensional unpressurised vesicles, or
self-avoiding loops of fixed length $N$,
behaves for large $N$ as $A_0\,N^{3/2}$, while their mean
square radius of gyration behaves as $R^2_0\,N^{3/2}$. The amplitude ratio
$A_0/R_0^2$ is computed exactly and found to equal $4\pi/5$.
The physics of the pressurised case, both in the inflated and collapsed
phases, may be usefully related to that of a complex $O(n)$ field
theory coupled to a $U(1)$ gauge field, in the limit $n\to0$.
\endtitlepage

\oneandahalfspace
Self-avoiding loops, or lattice polygons, have been studied extensively
as models for planar vesicles. In a pioneering paper,
Leibler, Singh and Fisher\refto{LSF} (LSF)
considered the statistics of the area and
shape of such loops, subject to an osmotic pressure difference
$\bar p$, and controlled by a rigidity parameter $\kappa$. While much of
the interesting physics arises as a result of the variation of this
latter quantity, these authors also observed interesting scaling
behaviour as a function of $\bar p$, when $\kappa=0$.
Specifically, they found, on the
basis of Monte Carlo studies and exact enumerations, that
for an ensemble in which the total length, or mass, $N$ of the loop is
fixed, the mean area and squared radius of gyration behave as
$$
\langle A\rangle_N\sim A_0N^{2\nu_A}Y(x),\qquad
\langle R^2_G\rangle_N\sim R_0^2N^{2\nu}X(x)   \eqno(a)
$$
where $x=\bar pN^{\varphi\nu}$, and $\nu=\frac34$ is the usual
self-avoiding walk exponent in two dimensions.\refto{NIEN}
LSF argued that $\varphi=2$, and conjectured $\nu_A=\nu$, a result
which was derived indirectly by Duplantier\refto{DUP} on the basis of
Coulomb gas arguments. These results were
confirmed and extended to measures of the shape dependence by Camacho
and Fisher,\refto{CF} and the lattice enumeration studies were carried
to higher orders in papers by Conway, Enting, Fisher, Guttmann and
Whittington.\refto{EG1,EG2,FGW,CEG}
One result of these studies\refto{CF,EG2,CG} was the
apparent universality of the ratio $A_0/R_0^2$.

This Letter describes an analytic approach to this problem.
Generalising the well-known correspondence of de Gennes,\refto{DG}
the problem of self-avoiding loops at fixed $\bar p$ and fixed
monomer fugacity $u$ (conjugate to $N$) is shown to be equivalent
to a complex $O(n)$ spin model coupled to a $U(1)$ gauge field.
This field-theoretic formulation of the problem immediately establishes
the scaling forms in \(a). Moreover, using the methods of
two-dimensional
conformal field theory and the Coulomb gas mappings of
Nienhuis,\refto{NIENDG} the ratio $A_0/R_0^2$ may be computed exactly
to be $4\pi/5$.

For the sake of definiteness, consider oriented
self-avoiding polygons on a
honeycomb lattice. The correspondence to a (complex) $O(n)$ spin
model\refto{DG,NIENDG} is as follows. Suppose that $s_a(r)$ label the
components ($a=1,\ldots,n$) of a complex-valued spin at the site $r$.
These spins are normalised so that $\tr
s_a^*(r)s_b(r')=\delta_{ab}\delta_{rr'}$ and $\tr s_a(r)s_b(r')=\tr
s_a^*(r)s_b^*(r')=0$. Then the partition function $Z\equiv\tr
\prod_{r,r'}\left(1+u\sum_as_a^*(r')s_a(r)+{\rm c.c.}\right)$, gives,
in the limit $n\to0$, the generating function for the number $p_N$
of (unoriented) self-avoiding loops per site: $Z=1+2n{\cal N}
\sum_Np_Nu^N+O(n^2)$, where $\cal N$ is the total number of sites.
Up to this point, the complex $O(n)$ model is completely equivalent
to the usual real $O(2n)$ model.
Consider now a unit current $J_\mu$ flowing along each link of a polygon
in the direction of its orientation.\refto{JM,CG,CM}
An explicit expression for this
current, when inserted into correlation functions of the lattice $O(n)$
model, is $J_\mu(r,r')=u(r'-r)_\mu\sum_a\left(s^*_a(r')s_a(r)-
{\rm c.c.}\right)$, where the lattice spacing has been taken equal to
unity.
In the continuum spin version of the $O(n)$ model,
in which the spins are replaced by a field $\Phi_a(r)$, $J_\mu$ is
just the $U(1)$ current, $(1/2i)\sum_a(\Phi_a^*\partial_\mu\Phi_a-
\Phi_a\partial_\mu\Phi_a^*)$, whose space integral generates the global
$U(1)$ symmetry $\Phi_a(r)\to e^{i\alpha}\Phi_a(r)$.
The area of a given loop is now given by
$$
A=-\frac12\int|r_1-r_1'|\delta(r_0-r_0')J_0(r)J_0(r')d^2rd^2r'
\eqno(b)
$$
introducing Cartesian co-ordinates $r=(r_0,r_1)$.
This expression
is valid for any non-self-intersecting loop (but not, in general,
for loops which
do self-intersect, since it weights different regions by the modulus of
the winding number of the loop around them.)
It differs from that for the {\it signed} area, which is proportional to
$\epsilon_{\mu\nu}\int r_\mu J_\nu(r)d^2r$. This latter quantity was
used in \ref{RUD} to study ordinary intersecting loops as a model of
vesicles. However, in the absence of any pressure difference its mean,
after averaging over orientations of the loop, vanishes identically,
and it is therefore not a suitable measure of the area.

In fact, \(b) is readily recognised as the expectation value of the polygon
regarded as a Wilson loop in a $U(1)$ gauge theory:
$$
A=\int\langle A_\mu(r)A_\nu(r')\rangle J_\mu(r)J_\nu(r')
d^2rd^2r'
\eqno()
$$
in the gauge $A_1=0$. It is straightforward to check that a similar
result holds also in a covariant gauge, where
$\langle A_\mu(r)A_\nu(r')\rangle=(1/2\pi)\bigg(
\delta_{\mu\nu}\ln|r-r'|-r_\mu r_\nu/r^2\bigg)$.
This is important,
since in this gauge the rotational invariance of the final result in
the continuum limit is manifest.

In the continuum limit near its critical point, the $O(n)$ lattice
model corresponds to a field theory with $O(n)$ symmetry,
which has been studied extensively in two dimensions.\refto{DF,ZAM,
CM} This continuum theory will possess a conserved $U(1)$ current
$J_\mu$. It then follows from the above discussion that the continuum
version of the generating function
for self-avoiding loops weighted by ${\rm e}^{\bar pA}$
is given by the $n\to0$
limit of an $O(n)$ theory, described by an action $S_0$,
coupled to an Abelian gauge field:
$$
Z=\tr\int{\cal D}A_\mu {\rm e}^{-S_0+ie\int J_\mu A_\mu d^2r-\frac14\int
F_{\mu\nu}^2d^2r}
\eqno(c)
$$
Integrating out the gauge field yields the identification
$\bar p=-e^2/2$. Several remarks may be made at this point. First,
since $A_\mu$ is dimensionless and $J_\mu$, being a conserved Noether current,
retains its canonical dimension of inverse length, it
follows that $\bar p$ has renormalisation group eigenvalue
$y_p=2$. From this, and simple renormalisation group scaling arguments,
follow the scaling laws in \(a) with $\nu_A=\nu$ and $\varphi=2$.
Second, the `physical' region of the $U(1)$ gauge theory, in which the
gauge couping $e$ is real and opposite charges attract, corresponds to
a negative internal osmotic pressure difference. In that region, LSF
find that, for large enough $N$, the loops collapse and behave like
branched polymers. This is to be expected from the field theory, since
in 1+1 dimensions a $U(1)$ gauge field provides a confining potential
so that the only asymptotic states are neutral. The world lines of
these bound states correspond to the filaments of the branched polymer.
The inflated phase, corresponding to $\bar p>0$, does not strictly
make sense in the field theory, since the vacuum would become unstable
to charge separation. As is well known,\refto{INST} the singularity in the free
energy for $e^2<0$ may be described in weak coupling by an instanton
calculation. For $\bar p>0$ it should be possible to neglect the
self-avoiding constraint, so that the action $S_0$ maybe replaced by
that of $n$ free complex scalar fields. In the first quantised picture,
the instanton configuration corresponds to a particle-antiparticle
pair being created at some imaginary time $r_0$ and annihilating at
time $r_0'$, their world lines describing a circle of radius $R$ (fixed
by extremising the total action) in euclidean space. This corresponds
exactly to the physical picture of an inflated vesicle.

Now return to the case $\bar p=0$, and the expression \(b) for the
mean area.
It is easier to work in the axial gauge, although the same
results are obtained in the covariant gauge.
Averaging \(b) over the ensemble of all self-avoiding loops,
$$
2n{\cal N}\sum_Np_N\langle A\rangle_Nu^N=-\frac12
\int|r_1-r_1'|\delta(r_0-r'_0)\langle
J_0(r)J_0(r')\rangle d^2rd^2r'
\eqno()
$$
where
$J$ is now the $U(1)$ current of the complex $O(n)$ theory,
in the limit $n\to0$.
Thus
$$
n\sum_Np_N\langle A\rangle_Nu^N=-\frac12
a_0\int_0^\infty r_1\langle J_0(r_1,0)J_0(0,0)
\rangle dr_1
\eqno(f)
$$
where $a_0$ is the area per site.
In general, by current conservation and dimensional analysis,
the correlation function $\langle J_\mu(r)J_\nu(0)\rangle$,
evaluated in the massive $O(n)$ field theory, has the form
$(\partial_\mu\partial_\nu-\delta_{\mu\nu}\partial^2)f(m|r|)$,
where $m$ is the  mass and
$f$ is a dimensionless scaling function, whose large $r$ asymptotic
behaviour may be evaluated
non-perturbatively using the form factor approach described in
\ref{CM}.
However,
at short distances, it becomes independent of $m$, and therefore
has the form\refto{JM} $k(n)(r_\mu
r_\nu-\frac12r^2\delta_{\mu\nu})/r^4$,
determined, up to the constant $k(n)$, by current conservation and
rotational symmetry. Since the normalisation of this current is fixed
by the requirement that its integral generate the $U(1)$ symmetry, the
number $k(n)$ is universal. Moreover, being a short-distance limit, it
should be calculable within the conformal field theory corresponding to
the massless complex $O(n)$ field theory. In field theory, $k$ is called
the chiral anomaly, since current conservation implies the existence
of a contact term, proportional to $\delta^2(r)$,
in the operator product $J_L(r)J_R(0)$ of the left- and and right-moving
currents $J_{L,R}=J_0\pm J_1$.
(For integral values of $n>1$
the $U(1)$ symmetry would be embedded in a Kac-Moody algebra and
$k(n)$ would be proportional to the
level number, but for $n<1$ such a concept does not appear to make
sense.)

It then follows that
the correlation function on the right hand side of \(f) behaves like
$-k(n)/2(r_1^2)$,
so the integral appears to diverge logarithmically at short distances.
In fact such a divergence must occur, since otherwise the integral
would be dimensionless and therefore independent of the mass $m$,
implying that the left hand side has no singularity as a function of
$u$. In fact, if $b$ is used as a short-distance cut-off, this
divergence must be of the form $\ln(mb)$. Using the fact that the mass
vanishes at the critical point $u=u_c$
according to $m\sim(u_c-u)^\nu$, it follows that the right hand side of
\(f) has the singular behaviour $-\frac14a_0k(n)\nu\ln(u_c-u)$ and hence
that, as $N\to\infty$,
$$
p_N\langle A\rangle_N\sim \frac14\sigma a_0 k'(0)\nu N^{-1}u_c^{-N}
\eqno(d)
$$
where the lattice-dependent integer $\sigma$ appears\refto{PR,CG} because,
on non-close
packed lattices, the series is in fact in $u^\sigma$ and therefore has
$\sigma$ equivalent
singularities on $|u|=u_c$.
Defining the amplitude $B$ by $p_N\sim BN^{-2\nu-1}u_c^{-N}$,
it follows that $BA_0=\frac14\sigma a_0 k'(0) \nu $,
which gives $\frac38k'(0)$ for the square lattice ($\sigma=2$, $a_0=1$).

The next step is to evaluate $k(n)$ using Coulomb gas
methods.\refto{NIENDG}
The mapping to the Coulomb gas proceeds in two stages. On the honeycomb
lattice, the expansion of the partition function of the complex $O(n)$
model yields a sum over configurations of non-intersecting oriented
loops, weighted by a factor of $u$ for each link and $n$ for each loop.
This latter factor may be written in a local fashion by incorporating
a factor $e^{\pm i\chi}$ at each occupied vertex,
depending on whether the oriented loop makes a turn through
$\pm\pi/3$ at that point. Thus (anti-)clockwise loops accumulate factors
of $e^{-6i\chi}$ or $e^{6i\chi}$ respectively.
After the summation over orientations, the appropriate factor of $2n$
per loop may be recovered by choosing $n=\cos6\chi$.
This model may then be mapped onto a solid-on-solid (SOS) model by
assigning heights $\phi(r)$ (which are
conventionally chosen to be integer multiples of $\pi$) to the sites of
the dual lattice.
Neighbouring heights on either side of an oriented bond differ by $\pm\pi$,
otherwise they are equal. By convention, the higher side is on the
right,
looking along the oriented bond.
This model is then supposed\refto{NIENDG} to renormalise,
in the long wavelength limit,
onto a Gaussian model with action $S_G=(g/4\pi)\int(\partial\phi)^2d^2r$,
where $g=2-6\chi/\pi$. However, there is a {\it caveat}: in the SOS model
the factors $e^{\pm i\chi}$ lead to the result $\langle
e^{-12i\chi\phi(r)/\pi}\rangle=1$, as may be seen
by direct calculation in the
fugacity expansion. In the Coulomb gas language, where $\phi$ is
interpreted as an electrostatic potential, this phenomenon corresponds
to a total electric charge $12\chi/\pi$ on the boundary, which
preserves overall neutrality. Thus all non-zero correlation functions
must
correspond to a total charge $-12\chi/\pi$ in the interior.

How should the $U(1)$ current of the complex $O(n)$ model be represented
in the SOS model? The naive candidate is simply $J^{SOS}_\mu\equiv(1/\pi)
\epsilon_{\mu\nu}\Delta_\nu\phi$, where $\Delta_\nu$ denotes a lattice
difference between sites of the dual lattice. This current is conserved
and has the property of taking the values $\pm1$ as required. But it
is nevertheless incorrect, since clockwise and anticlockwise loops
are counted with different phase factors, resulting in a net clockwise
current of $e^{6i\chi}-e^{-6i\chi}=2i\sin6\chi$ around each loop.
In addition, it may be seen that the correlation function
$\langle J^{SOS}_\mu(r)J^{SOS}_\nu(r')\rangle$
receives contributions when the links $r$ and
$r'$ are on different loops, a feature which is
absent in the $O(n)$ model. This correlation function also suffers from
having net charge zero, so the charge on the boundary is not cancelled.

However, there is another conserved current
\def\JT{\widetilde J}
$\JT^{SOS}_\mu\equiv\lambda\epsilon_{\mu\nu}\Delta_\nu
\left(e^{-12i\chi\phi/\pi}\right)$,
where the constant $\lambda$ is to be fixed. This has the property that
its expectation value around a given loop vanishes, as required, on
summing over both orientations, since this is proportional to
$(e^{-12i\chi}-1)e^{6i\chi}+(e^{12i\chi}-1)e^{-6i\chi}=0$.
Now consider the current-current correlation function, which is the
expectation value of
$J_\mu(r)J_\nu(r')$. For a given configuration in the
loop gas, after summing over both orientations,
this quantity takes the values $2n\hat r_\mu\hat r'_\nu$
if $r$ and $r'$ lie on
the same loop (where $\hat r$ is a unit vector along the link $r$),
and is zero if they lie on different loops.
A suitable candidate for this in the SOS model is therefore
$$
\langle\JT^{SOS}_\mu(r)J^{SOS}_\nu(r')\rangle=
(\lambda/\pi)\langle\epsilon_{\mu\gamma}\Delta_\gamma
e^{-12i\chi\phi(r)/\pi}\,\epsilon_{\nu\delta}\Delta_\delta\phi(r')
\rangle
\eqno(JTJ)
$$
since the quantity in $\langle\ldots\rangle$ is zero when $r$ and $r'$
are on different loops, for the same reason as above, and when they are
on the same loop, it takes the value
$\pi\hat r_\mu\hat r'_\nu
(e^{-12i\chi}-1)e^{6i\chi}-\pi(e^{12i\chi}-1)e^{-6i\chi}
=-4\pi i\hat r_\mu\hat r'_\nu\sin6\chi$.
Thus one should choose $\lambda=in/(2\sin6\chi)$.
It is somewhat curious that it is necessary to use different currents
$J$ and $\JT$ in this expression, but such a result is enforced by the
requirement that the total charge be $-12\chi/\pi$ for the
correlation function to be non-zero.

It is now straightforward to evaluate \(JTJ) in the Gaussian model,
replacing the lattice differences by derivatives, and using the
fact that $\partial_\mu\phi(r')e^{-12i\chi\phi(r)/\pi}
\sim(-12i\chi/g\pi)(r-r')_\mu(r-r')^{-2}e^{-12i\chi\phi(r)/\pi}$.
The result is of the form expected, with
$$
k(n)={12n\chi\over\pi^2g\sin6\chi}={2n\arccos n\over\pi\sqrt{1-n^2}
(2\pi-\arccos n)}
\eqno(KN)
$$
where $0\leq\arccos n\leq\pi$. For $n=1$, the model describes a single species
of charged boson with repulsive interactions, whose infrared behaviour is that
of
free fermions. In that case, one
finds $k=1/\pi^2$, as may be checked independently. For $n\to0$
\(KN) gives $k'(0)=2/3\pi$, so that $BA_0=\sigma a_0/8\pi$.

In order to eliminate the lattice-dependent factors from this otherwise
universal result, it may be combined with the relation
$BR^2_0=5\sigma a_0/32\pi^2$, which follows from a sum rule\refto{JC1}
which is a consequence of Zamolodchikov's $c$-theorem\refto{ZAMCTHM}.
This amplitude relation was first derived for the square lattice in
\ref{JC2}, and generalised in
\ref{CG}. The main result given in the introduction then follows.
The Table shows the comparison of the these predictions with results
of lattice enumerations and a Monte Carlo simulation of a continuum
model. The agreement is very satisfactory.

It is interesting to note that random loops with self-intersection,
for which $\langle R^2\rangle_N\sim N$, correspond to a free $O(n)$
theory, for which the current-current correlation function behaves
as $r^{-2}\ln r$ at short distances.\refto{JM} As a result, the mean
area of such walks (weighted as discussed earlier) behaves as
$N\ln N$ for large $N$. This is not surprising, since, as mentioned
above, such loops are weighted by their winding number, whose average
grows logarithmically.\refto{WIND}
In higher dimensions, \(f) may be generalised to relate the generating
function for the mean area of a loop {\it projected} onto a fixed plane
to a similar integral over the current-current correlation function.
Unfortunately, for $d>2$ the singular behaviour of this integral does
not come entirely from the short-distance behaviour, and therefore the
whole scaling function $f(mr)$ is required, rather than just
the coefficient
of its short-distance behaviour, which was calculated in an
$\epsilon$-expansion by Miller.\refto{JM}
However, this argument does imply that the mean projected area
grows as $N^{2\nu}$, in contradiction to the numerical findings
(with rather short series) of \ref{WANG}.

In this Letter I have shown how the theory of pressurised
two-dimensional vesicles without rigidity may be given a field-theoretic
basis. For zero pressure, this leads to an exact prediction for the
universal amplitude ratio $A_0/R^2_0$.
In the inflated phase $\bar p>0$, instanton techniques are
applicable and should yield an asymptotic expansion for the crossover
functions in \(a) for large argument $x$. This work is currently in
progress. The collapsed phase $\bar p<0$ corresponds to
confinement, and may provide an alternative field-theoretic way of describing
the so far only partially
solved problem of two-dimensional branched polymers.

I thank P.~Fendley, M.~E.~Fisher, A.~J.~Guttmann and J.~Miller
for correspondence and
discussions on this problem.
\vskip 1cm

\noindent{\bf Table.} Most accurate existing
estimates of the amplitudes $B$, $A_0$ and $R_0^2$
and their universal ratios compared with the predictions of this work.
\vskip 1cm

\input tables
\begintable
|Square|Triangular|Continuum|Prediction\crthick
$\sigma a_0$|2|$\surd3/2$|-|\crnorule
$B$|$0.5623^8$|$0.2640^6$|-|\crnorule
$A_0$|$0.1416^5$|$0.131^6$|$0.314\pm3^4$|\crnorule
$R_0^2$|$0.05631^{18}$|-|-|\cr
$BA_0/\sigma a_0$|0.03981|0.0399|-|${1\over8\pi}=0.0397887$\crnorule
$A_0/R_0^2$|2.515|-|$2.55\pm5^4$|${4\pi\over5}=2.51327$\endtable

\references

\refis{LSF} S.~Leibler, R.~Singh and M.~E.~Fisher, \journal
Phys. Rev. Lett., 59, 1989, 1987.

\refis{NIEN} B.~Nienhuis, \journal Phys. Rev. Lett., 49, 1062, 1982.

\refis{DUP} B.~Duplantier, \journal Phys. Rev. Lett., 64, 493, 1990.

\refis{CF} C.~J.~Camancho and M.~E.~Fisher, \journal
Phys. Rev. Lett., 65, 9, 1990.

\refis{EG1} I.~G.~Enting and A.~J.~Guttmann, \journal J. Stat. Phys.,
58, 475, 1990.

\refis{EG2} I.~G.~Enting and A.~J.~Guttmann, \journal J. Phys. A,
25, 2791, 1992.

\refis{FGW} M.~E.~Fisher, A.~J.~Guttmann and S.~G.~Whittington,
\journal J. Phys. A, 24, 3095, 1991.

\refis{CG} J.~L.~Cardy and A.~J.~Guttmann, \journal J. Phys. A, 26,
2485, 1993.

\refis{DG} P.~G.~de~Gennes, \journal Phys. Lett. A, 38, 339, 1972.

\refis{NIENDG} B.~Nienhuis, in {\sl Phase Transitions and Critical
Phenomena}, v. 11, C.~Domb and J.~L.~Lebowitz, eds. (Academic, 1986.)

\refis{JM} J.~Miller, \journal J. Stat. Phys., 63, 89, 1991.

\refis{RUD} G.~Gaspari, J.~Rudnick and A.~Beldjenna, \journal
J. Phys. A, 26, 1, 1993.

\refis{DF} Vl.~S.~Dotsenko and V.~A.~Fateev, \journal Nucl. Phys. B,
240, 312, 1984.

\refis{ZAM} A.~B.~Zamolodchikov, \journal Mod. Phys. Lett. A, 6, 1807,
1991.

\refis{CM} J.~L.~Cardy and G.~Mussardo, {\sl Nucl. Phys. B}, to
appear.

\refis{INST} This calculation is similar to that described in
J.~Schwinger, \journal Phys. Rev., 82, 664, 1951; and I.~Affleck,
O.~Alvarez and N.~S.~Manton, \journal Nucl. Phys. B, 197, 509, 1982;
although these actually refer to the case of an external field in
four dimensions.

\refis{JC1} J.~L.~Cardy, \journal Phys. Rev. Lett., 60, 2709, 1988.

\refis{ZAMCTHM} A.~B.~Zamolodchikov, \journal Zh. Eksp. Teor. Fiz.,
43, 565, 1986; [\journal JETP Lett., 43, 730, 1986.]

\refis{JC2} J.~L.~Cardy, \journal J. Phys. A, 21, L797, 1988.

\refis{WANG} J.~Wang, \journal J. Phys. A, 23, 4589, 1990.

\refis{PR} V.~Privman and S.~Redner, \journal J. Phys. A, 18, L781,
1985.

\refis{WIND} M.~E.~Fisher, V.~Privman and S.~Redner, \journal
J. Phys. A, 17, L569, 1984; B.~Duplantier and H.~Saleur, \journal
Phys. Rev. Lett., 60, 2343, 1988.

\refis{CEG} A.~R.~Conway, I.~G.~Enting and A.~J.~Guttmann, \journal
J. Phys. A, 26, 1519, 1993.

\endreferences

\endit